\documentclass[a4paper,fleqn]{cas-sc}
\usepackage{float}

\usepackage[numbers,comma, sort&compress]{natbib}
\usepackage[version=3]{mhchem}
\usepackage{amsmath,float}
 \usepackage{placeins} %%%%%
\def\tsc#1{\csdef{#1}{\textsc{\lowercase{#1}}\xspace}}
\tsc{WGM}
\tsc{QE}

\begin{document}
\let\WriteBookmarks\relax
\def\floatpagepagefraction{1}
\def\textpagefraction{.001}

\shorttitle{Software implementation for calculating the Chern and $\mathbb{Z}_{2}$ topological invariants with WIEN2k}    

\shortauthors{A. Gómez, O. Rubel}  

\title [mode = title]{Software implementation for calculating  Chern and \texorpdfstring{$\mathbb{Z}_{2}$}{Z2} topological invariants of crystalline solids with WIEN2k all-electron density functional  package}  

%\tnotemark[2]

%\tnotetext[2]{Hi} 


\author[1,2]{A. F. Gomez-Bastidas}[]

%\cormark[]

%\fnmark[]

\ead{gomezbaa@mcmaster.ca}

%\ead[url]{<URL>}

%\credit{<Credit authorship details>}

\affiliation[1]{organization={Department of Materials Science and Engineering, McMaster University},
            addressline={1280 Main Street West}, 
            city={Hamilton},
%          citysep={}, % Uncomment if no comma needed between city and postcode
            postcode={L8S 4L8}, 
            state={Ontario},
            country={Canada}}
\affiliation[2]{organization={School of Physics, Universidad Industrial de Santander},
            addressline={Carrera 27 Calle 09}, 
            city={Bucaramanga},
%          citysep={}, % Uncomment if no comma needed between city and postcode
            postcode={680002}, 
            state={Santander},
            country={Colombia}}

\author[1]{O. Rubel}[orcid=0000-0001-5104-5602]

% Footnote of the second author

% Email id of the second author
\ead{rubelo@mcmaster.ca}

\ead[url]{https://olegrubel.mcmaster.ca}



\cortext[1]{Corresponding author}


% \sep Topological phases \sep Topological insulator \sep Berry curvature
\begin{abstract}
We present two modules that expand functionalities of the all-electron full-potential density functional theory package \textsc{WIEN2k} for computation of the Chern and $\mathbb{Z}_{2}$ topological invariants. Characterization of topological properties relies on two methods: computing an evolution of hybrid Wannier charge centers for $\mathbb{Z}_{2}$ topological insulators (construction of maximally localized Wannier functions is not needed) and computing the Berry phase for a multitude of Wilson loops that discretize a 2D Brillouin zone for Chern insulators as well as for mapping the Berry curvature. The implementation is validated by testing on well-known materials that feature topologically non-trivial electronic states.
\end{abstract}

\maketitle

\section{Introduction}\label{sec:Intro}
Topological materials are new phases of matter outside the scope of Landau's theory of phase transitions that have generated considerable interest due to their unusual electronic transport properties \cite{Haldane_revmodphys_89_2017_10.1103/RevModPhys.89.040502}. Topological materials led to a paradigm shift in the theoretical description of condensed matter. It becomes necessary to expand the band theory of solids by introducing new concepts, namely the Berry phase and curvature \cite{BerryMV_procroyalsoclodn_392_1984,ZakJ_PhysRevLett_62_1989_10.1103/PhysRevLett.62.2747}, needed to explain observable physical phenomena, such as the quantum anomalous Hall (QAH) and quantum spin Hall (QSH) effects. Both topological phases (QAH and QSH) are associated with presence of an electronic bulk gap and gapless edge states \cite{Thouless_Physrevlett_49_1982_10.1103/PhysRevLett.49.405,Haldane_Physrevlett_61_1988_10.1103/PhysRevLett.61.2015}. The QAH phase is related to a single spin conducting channel at the edge, while there are two channels (with opposite spins and group velocities) for the QSH phase \cite{LiuXing_AnnualRevCondMatPhys_7_2016_10.1146/annurev-conmatphys-031115-011417}. These materials hold the potential for the development of innovative devices and applications in the emerging areas of quantum computing and spintronics \cite{HasanKane_RevModPhys_82_2010_10.1103/RevModPhys.82.3045,Ando_JournPhysSocJap_82_2013_10.7566/JPSJ.82.102001,QiLiang_RevModPhys_83_2011_10.1103/RevModPhys.83.1057}.

Topological phases can be characterized by a global property of their electronic structure, i.e., a topological invariant \cite{SimonDavid_IopConPhys_2018}. The QAH phase in 2D systems with broken time reversal symmetry (TRS) is characterized by the total Chern number ($C$) given by \cite{Thouless_Physrevlett_49_1982_10.1103/PhysRevLett.49.405}, \cite[chap.~3.2.2]{Vanderbilt_book} 
\begin{equation}\label{eq:Chernumber}
 C =\frac{1}{2 \pi}\sum_{n}^{\text{occ.}} \int_{\mathrm{BZ}}\mathbf{\Omega}^{(n)}(\mathbf{k}) \cdot d \mathbf{S}.
\end{equation}
Here the integral is taken over the whole Brillouin zone (BZ), $n$ is the band index (the summation runs over all occupied bands), and the integrand is a Berry curvature pseudovector\footnote{The Berry curvature is an antisymmetric tensor $\Omega_{\alpha\beta}^{(n)}(\mathbf{k})$ with only tree independent components (diagonal components are zero), which are traditionally converted to a pseudovector
$
    \mathbf{\Omega}^{(n)}(\mathbf{k}) = 
    \{ 
        \Omega_{yz}^{(n)}(\mathbf{k}), 
        \Omega_{zx}^{(n)}(\mathbf{k}), 
        \Omega_{xy}^{(n)}(\mathbf{k})
    \}
$
using Levi-Civita permutation symbol.}. It is defined as ${\mathbf{\Omega}^{(n)}}(\mathbf{k})=\nabla_{\mathbf{k}}\times \mathcal{A}^{(n)}(\mathbf{k})$, where $\mathbf{k}$ is an electron wave vector and $\mathcal{A}^{(n)}(\mathbf{k})$ is the Berry potential $\mathcal{A}^{(n)}(\mathbf{k})= i \langle u_{\mathbf{k}}^{(n)} | \nabla_{\mathbf{k}} u_{\mathbf{k}}^{(n)}\rangle$ with $|u_{\mathbf{k}}^{(n)}\rangle$ being a cell-periodic part of the Bloch function. The Chern number is an integer that determines the value of QAH conductivity as $\sigma_{\text{QAH}}= C\,e^{2}/h$ \cite{Thouless_Physrevlett_49_1982_10.1103/PhysRevLett.49.405}, where $e$ and $h$ are the charge of the electron and Planck's constant, respectively. 

The Berry curvature ${\Omega}(\mathbf{k})$ is not readily available in practical density functional theory (DFT) calculations. This limitation precludes straightforward integration of Eq.~\eqref{eq:Chernumber}. Instead, \citet{Soluyanov_PhysrevB_83_2011_10.1103/PhysRevB.83.235401} proposed to evaluate the Chern number by tracking the evolution of hybrid Wannier charge centers (HWCCs) \cite{Sgiarovello_PhysRevB.64.115202_64_2001_10.1103/PhysRevB.64.115202} from one boundary in the BZ ($k=0$) to its periodical image (e.g., $k=2\pi/a_i$)
\begin{equation}\label{eq:ChernPolarization}
    C= \sum_{v}^{\text{occ.}} w^{(v)}(2\pi/a_i) - \sum_{v}^{\text{occ.}} w^{(v)}(0) .
\end{equation}
Here $\mathbf{a}_i$ ($i=1,2,3$) are real space lattice vectors, $v$ is the Wannier function index (not to be confused with the band index, although the number of Wannier function for occupied states corresponds to the number of occupied bands). The HWCCs $w^{(v)}$ are calculated at discrete intermediate ($k=0 \ldots 2\pi/a_i$) points in the BZ to ensure a smooth evolution of $\sum_{v}w^{(v)}(k)$ between boundaries due to the gauge uncertainty of $2\pi$ in the phase. Fractional coordinates for HWCCs are implied.

For the QSH phase Kane, Mele, and Fu proposed a $\mathbb{Z}_{2}$ invariant present in TRS 2D or 3D materials \cite{Kane_Physrevlett_95_2005_10.1103/PhysRevLett.95.146802,Fu_Physrevlett_98_2007_10.1103/PhysRevLett.98.106803}. As a consequence of TRS, the Bloch states are divided into Kramer's pairs $|u_{\pm\mathbf{k},\alpha}^{I}\rangle$ and $|u_{\mp\mathbf{k},\alpha}^{II}\rangle$, where the $2N$ occupied bands are divided in $N$ pairs ($\alpha=1, \ldots, N$). Labels $I$ and $II$ distinguish the pairs such that under the TRS operation $\mathcal{T}$ they transform, up to a phase factor ($e^{i\theta_{k,\alpha}}$), as \cite{LiangKane_PhysrevB_74_2006_10.1103/PhysRevB.74.195312} 
\begin{align}\label{eq:krammers}
e^{i\theta_{k,\alpha}}\mathcal{T}|u_{\mathbf{k},\alpha}^{I}\rangle & = |u_{-\mathbf{k},\alpha}^{II}\rangle, \nonumber \\
-e^{i\theta_{-k,\alpha}}\mathcal{T}|u_{\mathbf{k},\alpha}^{II}\rangle & =|u_{-\mathbf{k},\alpha}^{I}\rangle.
\end{align}
$\mathcal{T}$ is defined in terms of the spin ($S$) and complex conjugation ($K$) operators as $\mathcal{T}=e^{i \pi S / \hbar} K$. Each class of pairs has a corresponding Chern number $C_{I}$ and $C_{II}$ of equal magnitude but opposite in sign ($C_{I}= -C_{II}$). Thus, the total $C = C_{I} + C_{II}$ vanishes, resulting in a null Hall conductance \cite{Kane_Physrevlett_95_2005_10.1103/PhysRevLett.95.146802}. Nevertheless, the $\mathbb{Z}_{2}$ invariant related to the Chern number as $\mathbb{Z}_{2} = |C_{I~\text{or}~II}| \mod 2$ allows us to distinguish between a trivial insulator and the QSH topological phase, which originates from the relativistic spin-orbit interaction. 2D materials are characterized by only one $\mathbb{Z}_{2}$ index; for the 3D case four independent indices are necessary \cite[chap.~5.3.1]{Vanderbilt_book}, \cite{Fu_Physrevlett_98_2007_10.1103/PhysRevLett.98.106803,Liang_Physrevb_76_2007_10.1103/PhysRevB.76.045302}
\begin{equation}\label{eq:Z2 in 3D}
    \mathbb{Z}_{2} = ( \nu_0; \nu'_1 \nu'_2 \nu'_3 ).
\end{equation}
The so-called `weak' indices $\nu'_1$,  $\nu'_2$, and $\nu'_3$ represent $|C_{I~\text{or}~II}| \mod 2$ invariants taken at three planes in reciprocal space cutting through time-reversal invariant momentum (TRIM) points at which $|u_{\mathbf{k},\alpha}^{I}\rangle=|u_{\mathbf{k},\alpha}^{II}\rangle$. The strong index $\nu_0$ differentiates between a weak and strong topological insulator \cite{Kane_Physrevlett_95_2005_10.1103/PhysRevLett.95.146802,Fu_Physrevlett_98_2007_10.1103/PhysRevLett.98.106803,Bercioux_Springinternpublis_2018_Topologicalmatter}. The latter implies that metallic surface states are present on \textit{any} exposed surface of the material \cite[chap.~5.3.3]{Vanderbilt_book}.

Various methodologies for determining the $\mathbb{Z}_{2}$ topological invariant have been proposed \cite{LiangKane_PhysrevB_74_2006_10.1103/PhysRevB.74.195312,Fukui_Physsocjapan_76_2007_10.1143/JPSJ.76.053702,Soluyanov_PhysrevB_83_2011_10.1103/PhysRevB.83.235401,Soluyanov_PhysRevB_83_2011_10.1103/PhysRevB.83.035108,Yu_Physrevb_84_2011_10.1103/PhysRevB.84.075119}. The calculation of the surface states from first principles can be done, but it is cumbersome due to a great computational cost associated with the need of a super-cell construction. For the case of materials presenting inversion symmetry (IS), knowledge of the parity eigenvalues at TRIM points in the BZ is utilized \cite{Kane_Physrevlett_95_2005_10.1103/PhysRevLett.95.146802}. Otherwise, for the general case of systems lacking this symmetry, a calculation can be done by integrating $\Omega(\mathbf{k})$ and $\mathcal{A}(\mathbf{k})$ over half of the Brillouin zone through an appropriate discretization method to avoid complications associated with the gauge fixing \cite{LiangKane_PhysrevB_74_2006_10.1103/PhysRevB.74.195312,Fukui_Physsocjapan_76_2007_10.1143/JPSJ.76.053702,Soluyanov_PhysRevB_83_2011_10.1103/PhysRevB.83.035108}. \citet{Soluyanov_PhysRevB_83_2011_10.1103/PhysRevB.83.035108} analyzed obstructions associated with construction of Wannier functions for topological materials, which evolved into a practical approach for computing the $\mathbb{Z}_{2}$ indices by tracking the evolution of HWCCs in planes cutting through TRIM points in reciprocal space \cite{Yu_Physrevb_84_2011_10.1103/PhysRevB.84.075119,Soluyanov_PhysrevB_83_2011_10.1103/PhysRevB.83.235401}. The latter approach builds upon the concept of time-reversal polarization introduced by \citet{LiangKane_PhysrevB_74_2006_10.1103/PhysRevB.74.195312} and became the most popular choice in \textit{ab initio} calculations. Topological invariants are related to a gapless character of HWCCs as the wave vector evolves between TRIM points \cite{Yu_Physrevb_84_2011_10.1103/PhysRevB.84.075119,Soluyanov_PhysrevB_83_2011_10.1103/PhysRevB.83.235401,Soluyanov_PhysRevB_83_2011_10.1103/PhysRevB.83.035108} (see section \ref{sec:Method} for more details).

\textsc{Z2Pack} \cite{Gresch_PhysRevB_95_2017_10.1103/PhysRevB.95.075146} and \textsc{WannierTools} \cite{QuanSheng_Compphyscomm_224_2018_10.1016/j.cpc.2017.09.033} are open-source packages which offer numerical implementations for computing the aforementioned topological invariants.  The \textsc{Z2Pack} accepts physical models based on an analytical Hamiltonian, a tight-binding Hamiltonian, or an input from \emph{ab initio} DFT codes, such as  \textsc{VASP}, \textsc{Quantum ESPRESSO}, and \textsc{ABINIT}. Those codes belong to the pseudopotential, plane-wave basis family. \textsc{WannierTools}, on the other hand, requires construction of a Wannier function tight-binding model based on the results obtained from an \emph{ab initio} code via the \textsc{Wannier90} package \cite{Pizzi_Physcondmatt_32_2020_10.1088/1361-648x/ab51ff}. To the best of our knowledge, only one numerical implementations for computing the $C$ and $\mathbb{Z}_{2}$ topological invariants based on the all-electron full-potential DFT code has been reported so far \cite{Wanxiang_Compphyscomm_183_2012_10.1016/j.cpc.2012.04.001}. Nonetheless, the program is not publicly available for use.

In this communication we introduce the implementation of two open-source programs that broaden the capability of the all-electron full-potential DFT package \textsc{WIEN2k} \cite{Blaha_Vienna,Blaha_JourChemPhys_7_2020_10.1063/1.5143061}  to the characterization of topological phases by computing both the $C$ and $\mathbb{Z}_{2}$ invariants. In addition to computing the topological invariants, this contribution allows us to generate a map of the Berry curvature $\Omega(\mathbf{k})$ in an arbitrary plane of the Brillouin zone, which can be employed for identifying regions with inhomogeneous $\Omega(\mathbf{k})$ values. The latter functionality complements the growing interest to direct experimental reconstruction of the Berry curvature in reciprocal space, which becomes a target for measuring an underlying topology of Bloch bands \cite{Mikitik_Physrevlett_82_1999_10.1103/PhysRevLett.82.2147,Liu_Physrevlett_107_2011_10.1103/PhysRevLett.107.166803,Mikitik_Physrevlett_93_2004_10.1103/PhysRevLett.93.106403,Mak_Science_344_2014_10.1126/science.1250140,Liu_Natphys_13_2017_https://doi.org/10.1038/nphys3946}. We employ the programs \textsc{Wien2wannier} \cite{Kunes_Compphyscomm_11_2010_10.1016/j.cpc.2010.08.005} and \textsc{BerryPI}  \cite{Rubel_Compphyscomm_184_2013_10.1016/j.cpc.2012.10.028}  for the calculation of overlap matrix elements and the Berry phase, respectively (both codes are implemented in \textsc{WIEN2k}). Our implementation does not require construction of Wannier functions thereby making characterization of topological materials more straightforward. This development complements the \textsc{WloopPHI} module \cite{Saini_Compphyscomm_270_2022_10.1016/j.cpc.2021.108147} for characterization of Weyl semimetals and completes the set of tools for computational characterization of topological materials in the \textsc{WIEN2k} DFT package. Finally, the program was validated by computing topological invariants of a well-established $\mathbb{Z}_2$ topological insulator \ce{Bi2Se3} and a predicted Chern insulator \ce{FeBr3}, as well as the Berry curvature map in monolayer \ce{MoS2}.

\section{Method}\label{sec:Method}
We consider the electronic ground-state for a periodic crystal, which is described by a single-particle mean-field Hamiltonian $H(\mathbf{k})$, which is a smooth function of the $\mathbf{k}$ crystalline wave vector. The eigenstates of this Hamiltonian are found through the solution of the Kohn-Sham equations given by the DFT \cite{HohenbergKohn_Physrev_136_1964_10.1103/PhysRev.136.B864,KohnSham_Physrev_140_1965_10.1103/PhysRev.140.A1133}.

\subsection{Hybrid Wannier charge centers}

\begin{figure}[h!]
    \centering
    \includegraphics{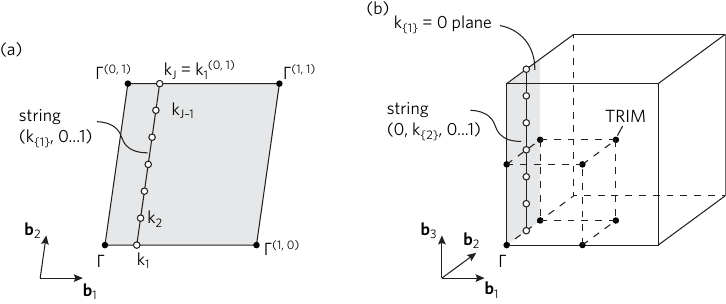}
    \caption{(a) String of $k$ points forming a Wilson loop in a 2D BZ used for calculation of HWCCs along direction of the second lattice vector. (b) The same construction in a 3D BZ for tracking of HWCCs along the third lattice vector and determination of $\mathbb{Z}_{2}$ indices. $k$ point coordinates are fractional in the basis of reciprocal lattice vectors $\mathbf{b}_i$ ($i=1,2,3$). The superscript indicates a periodic translation, e.g. $(0,1)$ implies translation along $\mathbf{b}_2$.}\label{fig:WCC-method}
\end{figure}

Our approach to computing HWCCs builds upon the method used in \textsc{Z2Pack} \cite{Gresch_PhysRevB_95_2017_10.1103/PhysRevB.95.075146}. Figure~\ref{fig:WCC-method}a shows the construction of a string of $k$ points that form a closed loop (Wilson loop) due to periodicity of the BZ (i.e., equivalent initial and final $k$ points $\mathbf{k}_1 \equiv \mathbf{k}_{J}$). The string is oriented along one of the reciprocal lattice vectors (the `Wannierization' direction). The Wilson loop is discretized with $J$ points.  All points on the string have a common $k_{i,\{1\}}$ coordinate (here $i$ is the point index, and the vector component index is in the curly brackets).

Overlap matrix elements are defined between the cell-periodic parts of two eigenstates as
\begin{equation}\label{eq:sec-2.1:S_mn}
    S_{mn}(\mathbf{k}_i,\mathbf{k}_j) = \langle u^{(m)}_{\mathbf{k}_i} | u^{(n)}_{\mathbf{k}_j} \rangle
    \quad m,n \in [B_\text{in},B_\text{fin}].
\end{equation}
Here $S$ is a square matrix with dimension given by the number of  bands in the interval of interest $[B_\text{in},B_\text{fin}]$. A cumulative overlap matrix for the Wilson loop of $k$ points is expressed as the matrix product
\begin{equation}\label{eq:sec-2.1:M cumulative}
    M = \prod_{i=1}^{J-1} S(\mathbf{k}_i,\mathbf{k}_{i+1}).
\end{equation}
Note the equivalency $\mathbf{k}_1 \equiv \mathbf{k}_{J}$ due to periodicity of the BZ (Fig.~\ref{fig:WCC-method}a). Eigenvalues of the cumulative overlap matrix
\begin{equation}\label{eq:sec-2.1:lambda}
    \lambda = \text{eig}(M)
\end{equation}
are used to obtain HWCCs (fractional coordinates) from a complex argument of each eigenvalue
\begin{equation}\label{eq:sec-2.1:w}
    w_{\{\alpha\}} = \frac{\arg(\lambda)}{2\pi} \mod 1.
\end{equation}
Here $\alpha \in \{1, 2, 3\}$ labels the direction of Wannierization (e.g., $w_{\{2\}}$ in Fig.~\ref{fig:WCC-method}a). It should not necessary be aligned with one of Cartesian directions ($x$, $y$, or $z$). Instead, the Wannierization direction is related to one of reciprocal lattice vector. For instance, if the Wannierization is performed along the first reciprocal lattice vector, the HWCCs will correspond to fractional coordinates along the first lattice vector in the real space. It should be emphasised that construction of maximally localized Wannier functions is \textit{not} required when computing HWCCs.

\subsection{Chern number}

\begin{figure}[h!]
    \centering
    \includegraphics{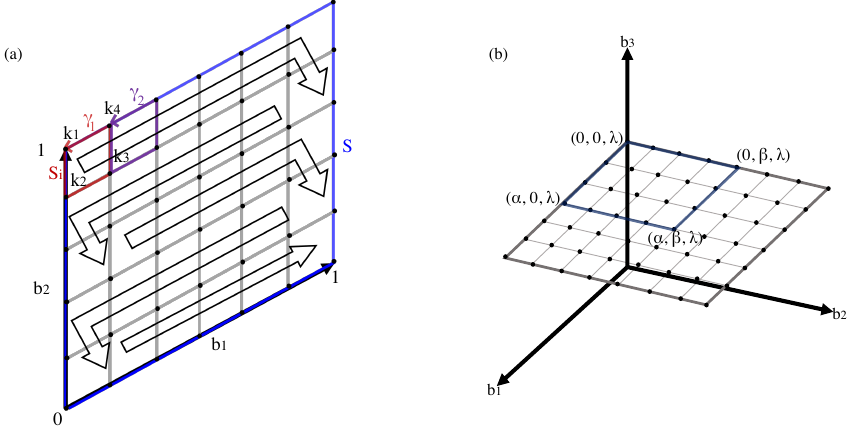}
      \caption{(a) Discretization of the BZ for the calculation of the Chern topological invariant. For each loop $S_{i}$ the Berry phase is calculated counterclockwise ($k_{1},..,k_{4},k_{1}$) and all the contributions are summed. The phase calculation direction consistency implies that the  contributions from segments shared between two loops cancel each other, and the Berry phase for the boundary $\partial S$ is obtained. One of the Berry phase unwrapping schemes (horizontal) for adjacent k point loops is presented. The unwrapping starts from $\gamma_2$ with respect to $\gamma_1$ and follows the direction of the black arrow (a vertical scheme is also performed and a continuity criterion is checked for both). (b) Boundary (light blue) discretization of the BZ (light gray) at the selected plane with height $\lambda$.}\label{fig:BZ-discret}
\end{figure}

For the calculation of the Chern number as per Eq.~\eqref{eq:Chernumber} it is necessary to perform an integration of the Berry curvature $\Omega(\mathbf{k})$ over the whole 2D BZ. In this context, the BZ is not taken as the Wigner-Seitz cell of the reciprocal space, but a primitive cell consisting of the parallelogram formed by the reciprocal lattice vectors as in Fig.~\ref{fig:BZ-discret}. This integration can be performed by subdividing the 2D BZ (\textbf{S}) into patches ($\mathbf{S}_i$) such that the total Chern number is given by the sum of Berry phases $\gamma^{(n)}_{\partial S_i}$ of occupied states accumulated on the boundary of each patch as originally proposed by \citet{Fukui_JournalPhysicsSocJap_74_2005__10.1143/JPSJ.74.1674}. The Berry phase is a $2\pi$ gauge-invariant quantity which represents a rotation in the complex phase of the cell-periodic part of the Bloch state $|u_{\mathbf{k}}^{(n)}\rangle$ as it traverses adiabatically a closed path over the BZ. It is defined in terms of the Berry potential $\mathcal{A}(\mathbf{k})$ and is related to the Berry curvature via Stoke's theorem as
\begin{equation}\label{eq:Berry phase}
\gamma^{(n)}_{\partial S}
=
\oint_{\partial S} \mathcal{A}^{(n)} (\mathbf{k}) \cdot d \mathbf{k}
=
\int_{S} {\mathbf{\Omega}^{(n)}} (\mathbf{k}) \cdot d \mathbf{S},
\end{equation}
where $\mathbf{S}$ is the vector of area normal to the surface, $S$ and $\partial S$ are the surface and its boundary, respectively. The calculation of the Chern number directly from `raw' $\gamma_{\partial S}$ values is hindered by the gauge uncertainty. For instance, if we were to compute the Berry phase on the whole boundary $\partial S$ of the BZ, it would not be possible to differentiate between the phase of 0 and $2\pi$, or equivalently between the Chern number of 0 and 1. Therefore, an alternative approach involves dividing $S$ into subspaces $S_i$  sufficiently small, so that the obtained change in phase between adjacent loops $\partial S_i$ is smooth enough in comparison to the gauge uncertainty of $2\pi$. Thus, for a patch $\mathbf{S}_i$ we have the Berry phase expressed in terms of HWCCs \eqref{eq:sec-2.1:w} computed on a Wilson loop (Fig.~\ref{fig:BZ-discret}a) comprised of discrete $k$ points \cite{Martin2020-fv}
\begin{equation}\label{eq:Patch_berryphase}
    \gamma_{\partial S_i} =
    2\pi \sum_v^{\text{occ.}} w^{(v)} \equiv
    \Phi_i.
\end{equation}
This quantity is also referred to as a flux of Berry curvature (or the Berry flux) $\Phi_i$ through the patch. For a discrete grid, the magnitude of the Berry curvature pseudovector projection onto a normal $\hat{\mathbf{n}}$ to the plane $S_i$ can be approximated as
\begin{equation}\label{eq:Berryflux}
        \mathbf{\Omega} (\mathbf{k}) \cdot \hat{\mathbf{n}}_{S_i} \approx \frac{\gamma_{\partial S_i}}{S_i}.
\end{equation}
Finally, the Chern number \eqref{eq:Chernumber} is computed as \cite{Martin2020-fv,Fukui_JournalPhysicsSocJap_74_2005__10.1143/JPSJ.74.1674,Palyi_Ashortcourseontopologicalinsulators_2016,Blanco_advquantmat_3_2020_10.1002/qute.201900117} 
\begin{equation}\label{eq:C_total}
    C = \frac{1}{2\pi} \sum_i^{\text{BZ}} \gamma_{\partial S_i}.
\end{equation}
%\ORadd{[please convey that $i$ should cover the whole $S$ to get $C$]} 
Since the Chern number is defined as the total flux of Berry curvature in the BZ, Eq.~\eqref{eq:C_total} can be viewed as a sum of locally calculated Berry fluxes through each patch, where the $i$ index runs over $\partial S_i$ in such way that the whole surface $\mathbf{S}$ is covered. The BZ constitutes a vectorial manifold, for the 2D case a $T^2$ torus, owing to the periodic boundary conditions of the crystal. Hence, the Chern number can be comprehended as the winding number of the Berry phase around this torus \cite{Bercioux_Springinternpublis_2018_Topologicalmatter}.

\section{Program Implementation}\label{sec:Implementation}

\subsection{Hybrid Wannier charge centers}\label{subsec:Imp_Z2}

It is implied that the user has converged an SCF calculation with SOC, and we continue to work in the same \texttt{case} directory. Calculation of HWCCs is managed by the \texttt{wcc.py} Python script, which is a part of the \textsc{BerryPI} package. The user needs to select the range of bands $[B_\text{in},B_\text{fin}]$, the Wannierization direction $\alpha$, the $k$ evolution direction and the range $k_{\text{evol}} \in [k_{\text{in}}, k_{\text{fin}}]$, and fix the $k$ point coordinate in the remaining direction. For example, in Fig.~\ref{fig:WCC-method}b the Wilson loop is constructed along the third direction, the HWCCs evolution is recorded along the second direction $k_{\text{evol}}\in[0,0.5]$, and $k_{\text{fix}}=0$ in the first direction. The user also selects the number of points $J$ on the Wilson loop and the number of discretization intervals along the $k$ evolution direction. Sensible numbers are about 10 and 20, respectively, to begin with.

The main loop is performed over the $k_{\text{evol}}$ variable. For each value, a Wilson loop is constructed (one at a time) and corresponding fractional coordinates of $k$ points are stored in the \texttt{case.klist} file in a native \textsc{WIEN2k} format as three integer numbers per $k$ point with a common integer divisor. It is important to note conventions used within \textsc{WIEN2k} to interpret $k$~point coordinates in the \texttt{case.klist} file \cite{Blaha_Vienna}. The BZ of a \textit{conventional} lattice is implied for F, B, CXY, CXZ, and CXZ orthorhombic lattices. The BZ of a \textit{primitive} lattice is used for other lattice types (P, H, R, CXZ monoclinic).

Once the \texttt{case.klist} file is set up, the script executes \textsc{BerryPI} to compute HWCCs for a given $k_{\text{evol}}$. Within \textsc{BerryPI}, the \textsc{WIEN2k} is called to compute wavefunctions for $k$ points on the Wilson loop followed by \textsc{Wien2wannier} to compute $S(\mathbf{k}_i,\mathbf{k}_{i+1})$ overlap matrices between adjacent $k$ point on the loop. The $S(\mathbf{k}_i,\mathbf{k}_{i+1})$  matrices are stored in the \texttt{case.mmn} file. Then, \textsc{BerryPI} reads the \texttt{case.mmn} file and computes HWCCs following Eqs.~\eqref{eq:sec-2.1:M cumulative}-\eqref{eq:sec-2.1:w}. The HWCCs $w_{\alpha}(k_{\text{evol}})$ are stored in a temporary \texttt{wcc\_i.csv} file, which concludes one cycle. Data on the evolution of HWCCs are consolidated in the \texttt{wcc.csv} file as a function of $k_{\text{evol}}$.

~

Sample of \texttt{wcc.csv} listing with $w^{(v)}$ values defined by Eq.~\eqref{eq:sec-2.1:w} for each Wilson loop stored in rows:
\begin{verbatim}
#k values are fractional coordinates in direction of the recip. latt. vec. G[2]
#WCC are evaluated on a Wilson loop in direction of the recip. latt. vec. G[3]
#k,WCC 1,WCC 2,WCC 3,WCC 4,WCC 5,WCC 6,WCC 7,...
0.000000,0.000000,0.000000,0.077836,0.077836,0.203946,0.203946,0.296585,...
0.026316,0.004659,0.004896,0.075512,0.078418,0.143604,0.233400,0.290768,...
0.052632,0.005981,0.013466,0.061571,0.080237,0.093847,0.246264,0.288668,...
0.078947,0.007861,0.020868,0.045604,0.079165,0.087893,0.251583,0.288239,...
0.105263,0.008441,0.023783,0.036129,0.077025,0.086447,0.257375,0.288128,...
0.131579,0.008095,0.023359,0.029160,0.075123,0.085686,0.264259,0.287875,...
0.157895,0.007344,0.021672,0.023798,0.073264,0.085003,0.270890,0.287507,...
...
\end{verbatim}

~

Sample of \texttt{wcc.py} screen output that shows evolution of the \textit{total} Berry phase $\sum_v w^{(v)}$ for each Wilson loop, which can be used to characterize Chern insulators:
\begin{verbatim}
Total Berry phase on each Wislon loop for bands 85-130:
------------------------------------------------------
 i           k            Phase wrap.    Phase unwrap.
                            (rad)           (rad)
------------------------------------------------------
 1  [***, 0.000, 0.000]     141.372           3.142
 2  [***, 0.026, 0.000]     141.544           3.314
 3  [***, 0.051, 0.000]     141.684           3.454
 4  [***, 0.077, 0.000]     141.780           3.550
...
39  [***, 0.974, 0.000]     147.482           9.252
40  [***, 1.000, 0.000]     147.655           9.425
------------------------------------------------------
Here "***" refer to the direction of the Wilson loop.
\end{verbatim}

Users have an option to enable a parallel \textsc{WIEN2k} calculation (it requires setting up a \texttt{.machines} file \cite{Blaha_Vienna}), add spin polarization, and an orbital potential (e.g., DFT+$U$). The SOC is enabled automatically.

\subsection{Chern Number}\label{subsec:Imp_C}

\textsc{CherN} is a \textsc{Python} script that divides a plane $S$ in the BZ into subsets $S_i$ and evaluates the Berry phase for each boundary $\partial S_i$ following the method described in section \ref{sec:Method} by recursively invoking the main \textsc{BerryPI} program. The script must be executed in the case directory after performing the standard \textsc{WIEN2k} self-consistent field calculation. For this purpose, the program receives as input: the band's range [$B_\text{in},B_\text{fin}$], the plane normal direction and height, the boundary range [$b_{1}^{\text{initial}},b_{1}^{\text{final}},b_{2}^{\text{initial}},b_{2}^{\text{final}}$] (if $b_3$ is selected as the plane normal direction) and finally, the discretization parameters $n_1$ and $n_2$. For example, Fig.~\ref{fig:BZ-discret}b presents a scenario for a plane in $b_3$ direction with height $\lambda$ and a boundary [$0,\alpha,0,\beta$]. It is worth noting that for the computation of the Chern number the boundary should cover the whole 2D BZ, i.e, $[0, 1, 0, 1]$, and the band range selected should be separated from other bands by an energy gap (isolated individual band analysis can be performed). Moreover, it is possible to switch the spin-polarized, parallel, and orbital potential (DFT+$U$) calculation flags. SOC is implied by default.

~

Sample of \texttt{CherN.py} input: 
\begin{verbatim}
    bands= [1,70]
    n_1 = 10
    n_2 = 10
    plane_dir = 3 
    plane_height = 0.0
    boundary = [0,1.0,0,1.0] 
    parallel = False  
    spinpolar = True 
    orbital = False 
\end{verbatim}

The script generates a mesh-grid discretization of $(n_1-1)\times(n_2-1)$ for the chosen boundary at a plane perpendicular to the selected direction and located at the constant height (see Fig.~\ref{fig:BZ-discret}b). Following this, the program generates the appropriate \texttt{case.klist} file for each loop $\partial S_i$ in the discretized grid and invokes the \textsc{BerryPI} code with the specified flags. As the Berry phase calculation inherently carries a $2\pi$ uncertainty, two phase unwrapping schemes are utilized (the python \texttt{numpy.unwrap} function) for all the calculated phases (see Fig.~\ref{fig:BZ-discret}b) to ensuring continuity of the Berry flux between adjacent patches. The acceptable criterion for continuity is $\Delta \gamma < \pi/2$ between any two adjacent patches. If the continuity criterion is not fulfilled, a warning message will be displayed at the end of the run. In the latter case it is \textit{imperative} to increase the mesh grid ($n_1,n_2$) until continuity is achieved. There is no \emph{a priori} way to estimate values of the discretization parameters. In general, materials with a smaller gap require a denser mesh. According to a Kubo formula for the Berry curvature based on the perturbation theory (written as a sum over all states \cite{Garg_AJP_78_2010}), the Berry curvature is inversely proportional to the square of the energy difference between states. Thus, the curvature is more localized in materials with a smaller gap, especially near to band crossings \cite{Yao_PhysRevLett_92_10.1103/PhysRevLett.92.037204,Stejskal_SciRep_12_10.1038/s41598-021-04076-z}. This fact should be taken as a marker for initial selection of the mesh density based on the band-structure analysis. Later we will give two examples of materials (\ce{MoS2} and \ce{FeBr3}) and encourage the reader to verify this trend.

Finally, the program calculates the total Chern number $C$ by employing Eq.~\eqref{eq:C_total} and stores the result, as well as the Berry curvature projection data (image matrix with values in units of rad~bohr$^2$ separated by comma) obtained from Eq.~\eqref{eq:Berryflux} in the \texttt{berrycurv.dat} file. This methodology differs from that chosen in the previously mentioned codes (\textsc{Z2Pack} and \textsc{WannierTools}) where Eq.~\eqref{eq:ChernPolarization} was employed, which implies the discretization of the Brillouin zone in a set of strings along the direction of one of the reciprocal lattice vectors. The two methods are equivalent from the Chern number perspective. In our case, however, the selection of loops also allow us to generate a map of the Berry curvature $\Omega(\mathbf{k})$ according to Eq.~\eqref{eq:Berryflux}. If the \textsc{Matplotlib} library is installed, a Berry curvature map is saved in \texttt{.png} and \texttt{.pdf} formats. 
The source code of \textsc{CherN} is available in the \textsc{BerryPI}  GitHub repository \cite{Rubel2022-berrypi}. The execution of \textsc{CherN} requires \textsc{WIEN2k} \cite{Blaha_JourChemPhys_7_2020_10.1063/1.5143061} and \textsc{BerryPI} \cite{Rubel_Compphyscomm_184_2013_10.1016/j.cpc.2012.10.028} (along with its dependencies) installed, as well as the \textsc{NumPy} library. 

\section{Validation}\label{sec:Validation}

\subsection{$\mathbb{Z}_{2}$ invariant in \ce{Bi2Se3} topological insulator}\label{subsec:Val_Bi2Se3}

\begin{figure}[h!]
    \centering
    \includegraphics{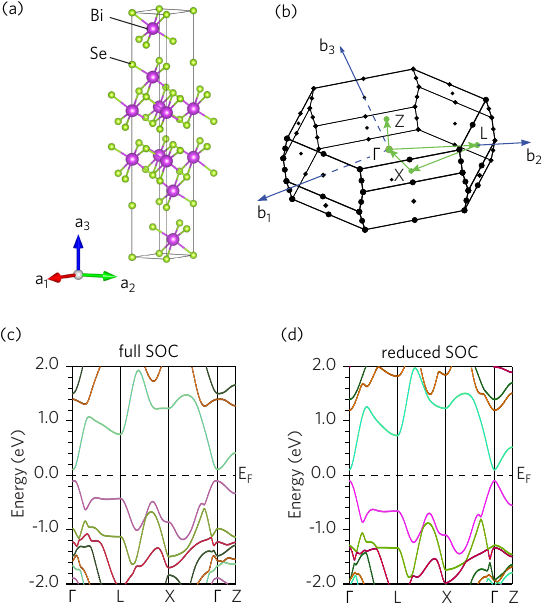}
    \caption{\ce{Bi2Se3}: (a) conventional unit cell, (b) primitive Brillouin zone with the $k$ path (green), (c) band structure with SOC (full strength, a non-trivial topological phase), (d) band structure with a reduced SOC strength (a trivial insulator).}\label{fig:Bi2Se3-band-struct}
\end{figure}

To validate the HWCCs implementation we compute invariants for \ce{Bi2Se3}, the well-known topological insulator \cite{Zhang_Natphys_5_2009_10.1038/nphys1270,Xia_natphys_5_2009_10.1038/nphys1274}. The structure of \ce{Bi2Se3} (space group 166, $R\bar{3}m$ from Ref.~\citenum{Nakajima_Jphyschemsol_24_1963_10.1016/0022-3697(63)90207-5}) is shown in Fig.~\ref{fig:Bi2Se3-band-struct}a (a conventional unit cell). The corresponding BZ (primitive cell) is depicted in Fig.~\ref{fig:Bi2Se3-band-struct}b. Figure~\ref{fig:Bi2Se3-band-struct}c presents the band structure of \ce{Bi2Se3} calculated in WIEN2k with SOC. It should be noted that SOC in WIEN2k is added perturbatively using a second variational method \cite{MacDonald_JPCSSP_13_1980}, and users need to be aware of the requirement to expand the basis set by adding empty states during the first scalar-relativistic step in \texttt{LAPW1} (the default is 5~Ry above the Fermi energy, but materials with strong SOC may require 10~Ry \cite{Kunes_PRB_64_2001,Blaha_Vienna}). (Calculation parameters are: Perdew-Burke-Ernzerhof (PBE) \cite{Perdew_physrevlett_77_1996_10.1103/PhysRevLett.77.3865} exchange-correlation functional; muffin tin radii $R_{\text{MT}}=2.5,2.47$~bohr for Bi and Se, respectively; 18 and 16 valence electrons for Bi and Se, respectively; $\min(R_{\text{MT}}) K_{\text{max}}=8$; $5 \times 5\times 5$ shifted k-mesh.) The obtained electronic structure corresponds to an insulator, but it is not possible to confirm the band inversion at $\Gamma$ due to SOC without an additional analysis.

\FloatBarrier

\begin{figure}[h!]
    \centering
    \includegraphics{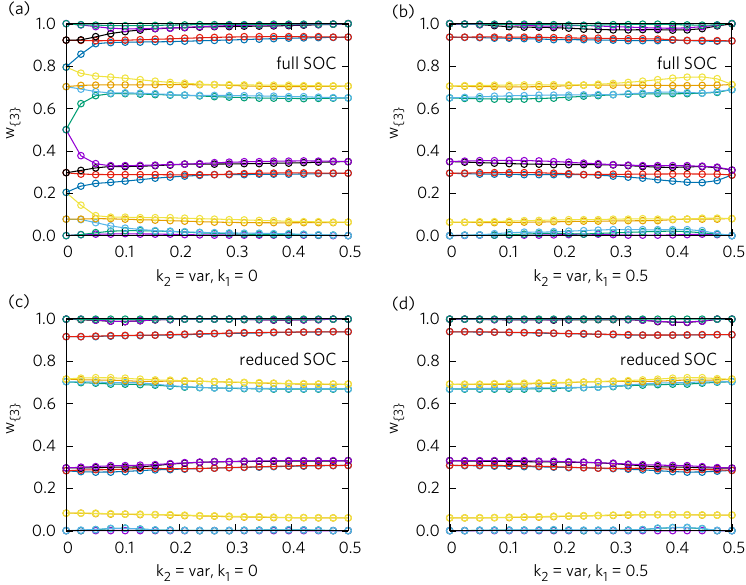}
    \caption{Evolution of HWCCs for \ce{Bi2Se3} in the BZ: (a,b) full SOC strength leads to a non-trivial topological insulator phase, (c,d) reduced SOC strength leads to a trivial insulator phase.  HWCCs are derived from a bundle of 18 bands below the Fermi energy. Coordinates of $k$ points vary between TRIM points. The Wannierization direction is given in brackets $\{\ldots\}$.}\label{fig:Bi2Se3-WCC-var-SOC}
\end{figure}

We investigate evolution of HWCCs between TRIM points using \texttt{BerryPI/wcc.py} in order to differentiate between a topologically trivial and non-trivial phase \cite{Yu_Physrevb_84_2011_10.1103/PhysRevB.84.075119,Soluyanov_PhysrevB_83_2011_10.1103/PhysRevB.83.235401}. For this purpose we selected a bundle of the 18 higher energy valence states. These states are well separated by a gap of ca.~3~eV from the remaining valence states. Results are shown in Fig.~\ref{fig:Bi2Se3-WCC-var-SOC}a,b. The panel (a) corresponds to HWCCs computed in a plane that starts at $\Gamma$ ($k_1=0, k_2=0$) and evolves towards L point ($k_1=0, k_2=0.5$), which is also one of TRIM points. The remaining dimension, i.e., the reciprocal lattice vector $\mathbf{b}_3$, is the Wannierization direction. The flow of HWCCs is \textit{gapless} on panel (a). The panel (b) shows an evolution of HWCCs between another set of TRIM points L ($k_1=0.5, k_2=0$) and X ($k_1=0.5, k_2=0.5$). The HWCCs evolution becomes \textit{gaped} when the plane passes away from $\Gamma$. We can conclude that the band inversion takes place at $\Gamma$ since the HWCCs evolution gap vanishes when approaching the center of the BZ indicating a non-trivial topological phase \cite{Soluyanov_PhysrevB_83_2011_10.1103/PhysRevB.83.235401,Yu_Physrevb_84_2011_10.1103/PhysRevB.84.075119}. This result is reminiscent of prior HWCCs calculations in \ce{Bi2Se3} (e.g., see Fig.~4 in Ref.~\citenum{Yu_Physrevb_84_2011_10.1103/PhysRevB.84.075119} and Fig.~7 in Ref.~\citenum{Gresch_PhysRevB_95_2017_10.1103/PhysRevB.95.075146}). Adding more occupied states to the analysis will not change the outcome, but can obscure the flow (see Fig.~4 in Ref.~\citenum{Soluyanov_PhysrevB_83_2011_10.1103/PhysRevB.83.235401} where 28 bands were used). It is worth mentioning that there is no exact mapping between band and HWCCs indices ($n \rightarrow v$) since more than one band generally contributes to one hybrid Wannier function. This information is contained in eigenvectors obtained when solving the eigenvalue problem \eqref{eq:sec-2.1:lambda}. However, these eigenvectors are presently neither stored nor analyzed.

\FloatBarrier

To show a contrast between the non-trivial and trivial insulator phases we artificially adjusted the SOC strength in WIEN2k (\texttt{LAPWSO} module) by increasing the speed of light $c$ from ca. 137 to 300~at.~units. Since the SOC strength is proportional to $c^{-2}$, the effective SOC was reduced by a factor of 4.8. At first glance, there are only minor visible changes to the band structure with reduced SOC (Fig.~\ref{fig:Bi2Se3-band-struct}d). However, the flow of HWCCs changes drastically (Fig.~\ref{fig:Bi2Se3-WCC-var-SOC}c,d). The evolution of HWCCs becomes gaped also in the plane passing through $\Gamma$ indicating that this change in the SOC strength tunes \ce{Bi2Se3} away from the topological insulator phase as also noticed by \citet{Zhang_Natphys_5_2009_10.1038/nphys1270}.

The gapless flow of HWCCs in an insulator (as in Fig.~\ref{fig:Bi2Se3-WCC-var-SOC}a) indicates winding of the HWCCs around the BZ, which is intimately connected with an adiabatic Thouless charge pumping in the bulk and leads to metallic edge states in a finite system  \cite{Soluyanov_PhysrevB_83_2011_10.1103/PhysRevB.83.235401,LiangKane_PhysrevB_74_2006_10.1103/PhysRevB.74.195312,Thouless_PhysRevB_27_1983_10.1103/PhysRevB.27.6083,Altshuler_science_283_1999_10.1126/science.283.5409.1864}. This bulk-boundary correspondence \cite{Hatsugai_PhysRevLett_71_1993_10.1103/PhysRevLett.71.3697} is captured by topological invariants. The full set of $\mathbb{Z}_{2}$ topological indices can be inferred from the flow of HWCCs between various TRIM points \cite{Soluyanov_PhysrevB_83_2011_10.1103/PhysRevB.83.235401}. In 3D
time-reversal-invariant insulators we need HWCCs for the following set \cite[chap.~5.3.1]{Vanderbilt_book}
\begin{align} 
k_1 = 0, k_2 = 0\ldots 0.5, w_{\{3\}} \quad &\text{and} \quad k_1 = 0.5, k_2 = 0\ldots 0.5, w_{\{3\}}, \label{eq:k-point-set-1}\\
k_1 = 0\ldots 0.5, k_2 = 0, w_{\{3\}} \quad &\text{and} \quad k_1 = 0\ldots 0.5, k_2 = 0.5, w_{\{3\}}, \label{eq:k-point-set-2}\\ 
k_2 = 0, k_3 = 0\ldots 0.5, w_{\{1\}} \quad &\text{and} \quad k_2 = 0.5, k_3 = 0\ldots 0.5, w_{\{1\}}. \label{eq:k-point-set-3}
\end{align}
Here the first (second) member in the set \eqref{eq:k-point-set-1} is related to the $\nu_1$ ($\nu'_1$) weak index. Panels (a) and (b) in Fig.~\ref{fig:Bi2Se3-WCC-var-SOC} correspond to the set \eqref{eq:k-point-set-1} leading to $\nu_1=1$ (gapless flow, one Wannier band is crossed when following the largest gap in the HWCC spectrum \cite{Soluyanov_PhysrevB_83_2011_10.1103/PhysRevB.83.235401} as a function of $k_2$) and $\nu'_1=0$ (gaped flow) indices, respectively. Results for the remaining two sets \eqref{eq:k-point-set-2} and \eqref{eq:k-point-set-3} are identical due to symmetry of the rhombohedral lattice and, therefore, are not shown. The remaining weak indices are $\nu_2=\nu_3=1$ and $\nu'_2=\nu'_3=0$. Since the flow of HWCCs is gapless in all directions originating at $\Gamma$, we should expect metallic states on \textit{any} surface plane, hence the strong topological insulator ($\nu_0=1$) \cite{Soluyanov_PhysrevB_83_2011_10.1103/PhysRevB.83.235401}. Following the definition \eqref{eq:Z2 in 3D}, the complete topological invariant of \ce{Bi2Se3} is $\mathbb{Z}_{2} = ( 1; 000 )$, which is consistent with numerous prior studies \cite{Zhang_Natphys_5_2009_10.1038/nphys1270,Soluyanov_PhysrevB_83_2011_10.1103/PhysRevB.83.235401,Yu_Physrevb_84_2011_10.1103/PhysRevB.84.075119}.

\subsection{Chern number of \ce{FeBr3}}\label{subsec:FeBr3}

2D magnetic materials are profiled as appropriate systems to host the AHC topological phase. The intrinsic magnetic ordering fulfills the time-reversal symmetry-breaking condition and, as a result, conductive edge states are expected to appear in materials with non-trivial Chern numbers. Therefore, we validate our \textsc{CherN} module by employing it on the \ce{FeBr3} monolayer, reported as a hypothetical Chern insulator with a topological invariant value $|C| = 1$ \cite{Olsen_Physrevmat_3_2019_10.1103/PhysRevMaterials.3.024005,Li_Physchemphsy_21_2019_10.1039/C8CP07781A,Zhang_arXiv_2017_https://doi.org/10.48550/arxiv.1706.08943}.%\ORadd{[would it be correct to say that the material is `hypothetical'? to give the reader a sense of feasibility for experimental validation]}

First, to obtain the \ce{FeBr3} monolayer structure we performed the ionic relaxation employing the \textsc{VASP} DFT package \cite{Kresse_physrevb_47_1993_10.1103/PhysRevB.47.558,Kresse_compmatscience_6_1996_10.1016/0927-0256(96)00008-0}. The projector augmented wave pseudopotential (PAW) method \cite{Bloch_physrevb_50_1994_10.1103/PhysRevB.50.17953,Kresse_physrevb_59_1999_10.1103/PhysRevB.59.1758} was utilized with cut-off energy of 500~eV for the plane-wave basis. Pseudopoltentials with 8 and 7 valence electrons were selected for Fe and Br, respecively. The exchange and correlation functional used was the Perdew-Burke-Ernzerhof (PBE) \cite{Perdew_physrevlett_77_1996_10.1103/PhysRevLett.77.3865} and the $\Gamma$-centered Monkhorst-Pack k-mesh grid \cite{Monkhorst_physrevb_13_1976_10.1103/PhysRevB.13.5188} of $13\times 13\times 1$ was selected. Both the cell parameters and internal atomic positions were fully relaxed with a force criterion of less than $10^{-2}$~eV~$\text{\AA}^{-1}$ on all atoms. Additionally, a vacuum of 20~{\AA} was set to avoid inter-layer coupling. The obtained structure belongs to the $P\overline{3}$1m (162) space group and the optimized lattice parameter of $a=6.297~\text{\AA}$ is in good agreement with Ref.~\citenum{Zhang_arXiv_2017_https://doi.org/10.48550/arxiv.1706.08943}.

\begin{figure}[h!]
    \centering
    \includegraphics{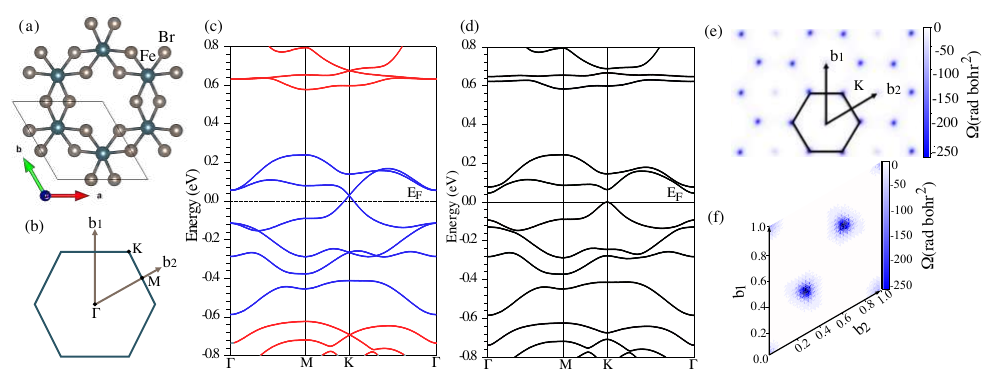}
    \caption{Monolayer \ce{FeBr3}: (a)  crystal structure, (b) BZ and high-symmetry points, (c,d) electronic band structure (spin-up:red, spin-down:blue) without and with SOC, respectively, (e) map of the Berry curvature pseudovector component $\Omega_z(\mathbf{k})$ in reciprocal space, (f) distribution of the Berry curvature pseudovector component $\Omega_z(\mathbf{k})$ in the primitive BZ (fractional coordinates).}\label{fig:FeBr3}
\end{figure}
\FloatBarrier

The obtained structure was verified with the \textsc{WIEN2k} package to ensure that the atomic forces do not exceed 2~mRy~bohr$^{-1}$ (ca.~0.05~eV~{\AA}$^{-1}$). Following this, the self-consistent field electronic structure calculation was performed within the spin-polarized and second-variational SOC framework, a parameter $\min(R_\text{MT})K_\text{max}=7$, and a k-point mesh grid of $15 \times 15 \times 1$ were taken. Here $K_\text{max}$ is the largest reciprocal-lattice vector size (plane wave cut-off), and atomic sphere radii are $R_\text{MT}$= 2.26, 2.15~bohr for Fe and Br, respectively. The energy cut-off separating the core from valence states was such that 14 and 17 electrons were considered as valence for Fe and Br, respectively. The calculation was initialized with a magnetic moment value of 1$\mu_{B}$ per formula unit, with Fe as the magnetically active atom, and the ground state found was ferromagnetic with a magnetic moment of 0.99$\mu_{B}$ per formula unit, henceforth, breaking the time-reversal symmetry of the system. Figure~\ref{fig:FeBr3}a,b shows the \ce{FeBr3} monolayer structure and the BZ, respectively. In Fig.~\ref{fig:FeBr3}c the electronic bandstructure without SOC is presented, where a crossing at the Fermi level (between bands 130 and 131) can be observed at the $K$ high-symmetry point. A SOC-induced gap opening of $47$~meV can be evidenced in Fig.~\ref{fig:FeBr3}d, which lifts the degeneracy at $E_{\text{F}}$ and may suggest the Chern insulator phase \cite{Olsen_Physrevmat_3_2019_10.1103/PhysRevMaterials.3.024005,Li_Physchemphsy_21_2019_10.1039/C8CP07781A,Zhang_arXiv_2017_https://doi.org/10.48550/arxiv.1706.08943}.

The projected Berry curvature map for \ce{FeBr3} is presented in 
Fig.~\ref{fig:FeBr3}e along with the detailed map in the primitive BZ (fractional coordinates) shown in Fig.~\ref{fig:FeBr3}f. It can be observed that the source of Berry curvature comes from the high symmetry point $K$  where the SOC gap emerged. The calculation was performed with the spin polar and parallel flags, the range of occupied bands $[1,130]$ was selected, and the discretization parameters $n_1 = n_2 = 33$ were used. It should be noted that for obtaining reliable results these parameters must be increased until the continuity criterion is achieved, in this case for values higher than $n_1 = n_2 = 21$ (higher values were employed for better graph resolution). Moreover, the lattice vector perpendicular to the monolayer ($\textbf{a}_3$) was selected as the \texttt{plane\_dir} parameter, and the plane height was set to 0.  The obtained Chern number given by Eq.~\eqref{eq:C_total} is $|C|=1$, which agrees with prior theoretical studies \cite{Olsen_Physrevmat_3_2019_10.1103/PhysRevMaterials.3.024005,Li_Physchemphsy_21_2019_10.1039/C8CP07781A,Zhang_arXiv_2017_https://doi.org/10.48550/arxiv.1706.08943}.
%\ORadd{[perhaps you can comment of how did you arrive to this choice of $n$? what happens if $n$ is small?]}

\subsection{Berry curvature of \ce{MoS2}}\label{subsec:Val_MoS2 monolayer}

\begin{figure}[h!]
    \centering
    \includegraphics{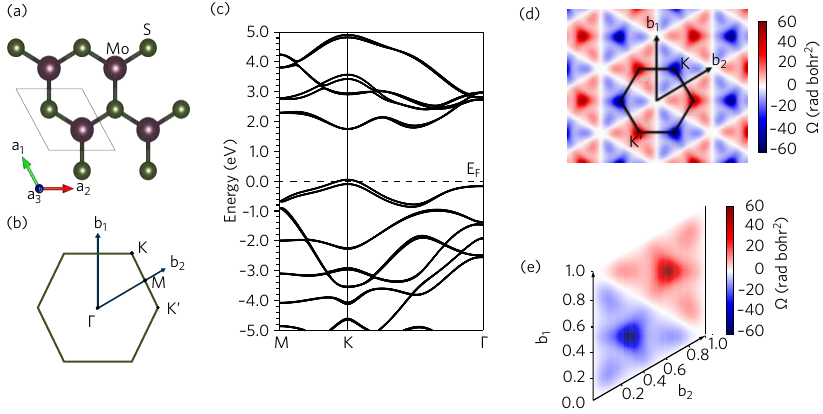}
    \caption{Monolayer \ce{MoS2}: (a) crystal structure, (b) BZ, (c) calculated electronic band structure with SOC, (d) map of the Berry curvature pseudovector component $\Omega_z(\mathbf{k})$ in reciprocal space, (e) distribution of the Berry curvature pseudovector component $\Omega_z(\mathbf{k})$ in the primitive BZ (fractional coordinates).}\label{fig-MoS2}
\end{figure}

For the validation of the Berry curvature maps, the \ce{MoS2} monolayer was selected due to its characteristic behaviour, namely, the opposite Berry curvature at the corners of the hexagonal Brillouin zone ($K$ and $K'$ high symmetry points as in Fig.~\ref{fig-MoS2}b) \cite{Feng_PhysRevB_86_2012_10.1103/PhysRevB.86.165108,Xiao_PhysRevLett_108_2012_10.1103/PhysRevLett.108.196802}. For this reason this material has attracted attention due to the potential uses in the emerging field of valleytronics  \cite{Xiao_PhysRevLett_108_2012_10.1103/PhysRevLett.108.196802,Lembke_AccChemRes_48_2015_10.1021/ar500274q,Wang_Envscietech_51_2017_10.1021/acs.est.7b01466,Li_accchemres_47_2014_10.1021/ar4002312}. Nevertheless, as this is a non-magnetic monolayer the TRS breaking condition is not met, and it cannot be a Chern insulator ($C = 0$).

\FloatBarrier

The self-consistent field calculations were carried out with the full-potential linearized augmented plane-wave DFT method implemented in \textsc{WIEN2k}. The experimental data from Ref.~\citenum{MoS2_exp} were utilized for the crystal structure, fixing the lattice parameter of the monolayer to $a_1=3.16~\text{\AA}$, which belongs to the $P \overline{6} m 2$ (187) space group, as presented in Fig.~\ref{fig-MoS2}a. To avoid interlayer coupling the distance between layers (center to center) was set at $a_3=12.3$~{\AA}, which corresponds to the double value of interlayer distance in bulk. The PBE generalized gradient approximation \cite{Perdew_physrevlett_77_1996_10.1103/PhysRevLett.77.3865} was selected for the exchange and correlation functional. For the electronic ground states the parameter $\min(R_{\text{MT}}) K_{\text{max}}=7$ was utilized (muffin tin radii $R_{\text{MT}}=2.44,2.10$~bohr for Mo and S, respectively) with a k-mesh grid of $16 \times 16 \times 1$. The energy cut-off separating the core from valence states was such that 14 and 6 electrons were considered as valence for Mo and S, respectively; the charge leakage was checked. The spin-orbit coupling was taken into account.

The calculated ground state was found to be nonmagnetic (resulting in $|C|=0$), and the electronic band structure is presented in Fig.~\ref{fig-MoS2}c. Here it can be seen that monolayer \ce{MoS2} is a semiconductor with a direct band gap of 1.8~eV \cite{Xiao_PhysRevLett_108_2012_10.1103/PhysRevLett.108.196802}. For calculation of the Berry curvature map, the k-parallel flag was employed, the selected range of occupied bands was $[1,26]$, the selected \texttt{plane\_dir} parameter corresponded to the lattice vector perpendicular to the monolayer (i.e., $\textbf{a}_3$), and the plane height was set to 0. The discretization parameters $n_1 = n_2 = 51$ were selected, while the continuity of Berry curvature was achieved starting from  $n_1 = n_2 = 32$. The inherent inversion symmetry breaking due to the monolayer configuration combined with the strong SOC result in appearance of opposite sign contributions to the Berry curvature with peaks at the $K$ and $K'$ high symmetry k-points. Moreover, $\Omega(\mathbf{k})$ vanishes at the $\Gamma$ and $M$ high symmetry points. These features of the Berry curvature map are in good agreement with prior theoretical studies by \citet{Feng_PhysRevB_86_2012_10.1103/PhysRevB.86.165108}.

\textbf{Data availability}: All structure files, WIEN2k workflow and input sections for \texttt{wcc.py} and \texttt{CherN.py} required to reproduce results reported here are freely available from Zenodo repository \cite{Zenodo_10.5281/zenodo.7761198}.

\section{Conclusion}\label{sec:Conclusions}

We introduced two software modules for computing topological invariants from the density functional theory framework implemented in the all-electron full-potential package \textsc{WIEN2k}. Characterization of topological properties relies on computing an evolution of hybrid Wannier charge centers (for $\mathbb{Z}_{2}$ topological insulators) and Berry phase for a multitude of Wilson loops that discretize a 2D Brillouin zone (for Chern insulators and the Berry curvature map). Validation of the program was confirmed by testing on the well-known materials with topological features: \ce{Bi2Se3} as a $\mathbb{Z}_{2}$ topological insulator, monolayer \ce{FeBr3} as a Chern insulator, and monolayer \ce{MoS2} with a peculiar Berry curvature map. The acquired results agree with the available experimental and computational data. The analysis of topological characteristics can be performed directly after completing a self-consistent field calculation without constructing maximally localized Wannier functions. This feature makes these computational tools attractive for the study and prediction of topological materials especially in a high-throughput computational screening environment. 

\section{Acknowledgements}\label{sec:Acknowledgements}
We thank to Peter Blaha (TU Vienna) for reading the manuscript and giving us constructive feedback. A.F.G.-B. would like to thank the MITACS organization for the financial support and logistics provided through the Globalink program. Calculations were performed using the Compute Canada infrastructure supported by the Canada Foundation for Innovation under John R. Evans Leaders Fund.


%\bibliographystyle{cas-model2-names}
\bibliographystyle{unsrtnat}

%\ORadd{[it seems that all refs. are missing the page number for articles]}

\bibliography{bibliography}

\bio{}
\endbio


% Use the style of numbering in square brackets.
% If nothing is used, default style will be taken.
%\begin{enumerate}[a)]
%\item 
%\item 
%\item 
%\end{enumerate}  

%\begin{itemize}
%\item 
%\item 
%\item 
%\end{itemize}  

%\begin{description}
%\item[]
%\item[] 
%\item[] 
%\end{description}  

% Figure
%\begin{figure}[<options>]
%	\centering
%		\includegraphics[<options>]{}
%	  \caption{}\label{fig1}
%\end{figure}

%\begin{table}[<options>]
%\caption{}\label{tbl1}
%\begin{tabular*}{\tblwidth}{@{}LL@{}}
%\toprule
%  &  \\ % Table header row
%\midrule
% & \\
% & \\
% & \\
% & \\
%\bottomrule
%\end{tabular*}
%\end{table}

% Uncomment and use as the case may be
%\begin{theorem} 
%\end{theorem}

% Uncomment and use as the case may be
%\begin{lemma} 
%\end{lemma}

%\appendix

%\item 
%\item 
%\item 
% Keywords
\end{document}